%% file: paper.tex
\documentclass[]{fairmeta}
\usepackage{amsmath}
\usepackage{amssymb}
\usepackage{url}

\title{EmoRES-TTS: Residual-Enhanced Vector \\Steering for Emotional Speech Generation}

\author[1,3,*]{Kuan-Po Huang}
\author[2]{Haohe Liu}
\author[2,]{Puyuan Peng}
\author[1,]{Haibin Wu}
\author[1,]{Zhaoheng Ni}
\author[3,]{Hung-yi Lee}
\author[1,]{Jinwon Lee}
\author[1,]{Neha Chachra}

\affiliation[1]{Reality Labs at Meta}
\affiliation[2]{FAIR at Meta}
\affiliation[3]{National Taiwan University}

\contribution[*]{Work done at Meta}

\abstract{Emotion-conditioned text-to-speech (TTS) models may fail to express the requested emotion reliably, and improving controllability by additional training is costly in both computation and emotion-labeled speech training data. We therefore study vector steering, a training-free approach that modifies the internal representations of a frozen model. CoCoEmo, a conventional vector steering method for emotion TTS, treats each emotion vector as an indivisible direction controlled by a single global strength, limiting adherence to the requested emotion. In this work, we first discover that an emotion vector can be decomposed into a shared component that moves speech away from neutral expression and a residual component that directs generation toward the requested emotion. Building on this finding, we propose Emotion Residual-Enhanced Steering for TTS (EmoRES), a novel method that controls the two components without retraining the backbone. On IEMOCAP, EmoRES outperforms CoCoEmo across all four objective emotion metrics on the IndexTTS-2 and CosyVoice2 backbones. Rank correlation improves by 26.13 and 12.97 percentage points, corresponding to relative gains of 118.8\% and 33.1\%, while emotion hit rate improves by 12.95 and 6.92 points, corresponding to relative gains of 20.1\% and 9.8\%. Human evaluation further shows a relative improvement up to 35.0\% in the rate at which listeners correctly identified the dominant requested emotion and up to a 17.3\% improvement in fidelity, while listeners prefer EmoRES for naturalness in up to 63.8\% of pairwise comparisons. Component ablations further demonstrate that effective control benefits from preserving the shared component while strengthening the residual of the emotion steering vectors.}

\date{September 2026}

\metadata[Code]{\url{https://github.com/facebookresearch/EmoRES-TTS}}

\begin{document}

\maketitle

\section{Introduction}
\label{section:intro}

\input{intro}

\section{Related Work}
\label{section:related_work}

\input{related_work}

\section{Method}
\label{section:method}

\input{method}

\section{Experimental Setup}
\label{section:exp_setup}

\input{exp_setup}

\section{Results and Discussion}
\label{section:results}

\input{results}

\section{Conclusions and Future Work}
\label{section:conclusions}

\input{conclusions}

\clearpage
\bibliographystyle{assets/plainnat}
\bibliography{paper}

\clearpage
\beginappendix

\section{Geometric illustration of EmoRES}
\label{sec:app-geometry}
\input{appendix/EmoRES_illustration}

\section{Metrics}
\label{app:metrics}
\input{appendix/metrics}

\section{Experimental details}
\label{sec:app-exp-detail}
\input{appendix/exp_details}

\clearpage
\section{Component ablation and random-direction controls}
\label{sec:app-ablmix}
\input{appendix/rand}

\newpage
\section{Steering with emotional speech prompts}
\label{sec:app-single-emo-prompt}

\input{appendix/emoprompt_single}

\clearpage
\section{Steering Without Native Emotion Conditioning}
\label{sec:app-no-conditioning}
\input{appendix/nocond}

\clearpage
\section{Different speech emotion recognizers}
\label{sec:app-3-ser}
\input{appendix/threejudge_iemocap}

\end{document}

%% file: intro.tex
Emotional expression is an important dimension of speech synthesis~\citep{triantafyllopoulos2023overview}. Since the same sentence may be spoken with different emotions, TTS systems need to control emotional expression while producing accurate linguistic content. This capability is important for conversational agents~\citep{chiba2018effect}, narration~\citep{liu2024storytts}, accessibility~\citep{fiannaca2018voicesetting}, and dubbing~\citep{cong2025emodubber}, where the requested emotion must be conveyed without compromising too much naturalness, intelligibility, or speaker identity~\citep{xie2025emosteer}.

Existing emotional TTS systems commonly provide control through native conditioning inputs, such as natural-language instructions~\citep{guo2023prompttts,yang2025emovoice,cosyvoice2} or emotion embeddings~\citep{indextts2}. These interfaces rely on emotion-aware training or specialized conditioning modules. However, we observe that native emotion conditioning alone does not always generate speech that strongly matches the requested emotion.

Vector steering~\citep{subramani2022extracting}, a form of activation steering, is an alternative that does not require retraining the model or relying solely on the model's native conditioning interface. It identifies directions associated with desired behaviors in a model's internal representations and adds them to the activations during inference, while leaving the model parameters unchanged \citep{turner2023steering,zou2023representation,rimsky2024steering}. Recent work extends this approach to emotional TTS by extracting directions from the activation differences between emotional and neutral speech \citep{xie2025emosteer,cocoemo}. These directions can strengthen emotional expression when the embedding or instruction conditioning produces a weak response, while a continuous steering strength controls the steering magnitude.
CoCoEmo~\citep{cocoemo} establishes this approach for modern language-model (LM)-based TTS systems~\citep{indextts2,cosyvoice2}. It extracts mean-difference emotion vectors from selected speech LM layers and injects the requested direction at inference time. However, CoCoEmo treats each emotion vector as a single unit governed by one global steering strength. This leaves the structure shared across the emotion vectors unexamined.

In our work, we observe that every categorical emotion vector contains two components: a shared component pointing from the neutral mean toward the centroid of emotional activations, and a residual component pointing from the centroid toward the requested emotion. Conventional steering scales these two components together. This is restrictive because moving speech away from neutral expression and directing it toward a particular emotion need not require the same strength. When stronger category-specific control is needed, increasing the global steering strength also amplifies the shared component. Conventional steering therefore cannot adjust the relative contributions of the two components, which may be suboptimal when they require different strengths.

In this work, we hypothesize that the shared component primarily moves generated speech away from neutral expression, whereas the residual directs generation toward the requested emotion. Based on this hypothesis, we propose \underline{Emo}tion \underline{R}esidual-\underline{E}nhanced \underline{S}teering for TTS, or EmoRES-TTS (abbreviated as EmoRES), a novel training-free method that controls the two components independently and enhances the residual steering strength relative to the shared component.
In Section~\ref{sec:main-results}, we evaluate EmoRES on IndexTTS-2~\citep{indextts2} and CosyVoice2~\citep{cosyvoice2}, which use different native emotion-conditioning interfaces. On IEMOCAP, EmoRES strengthens the correlation between the requested emotion proportions and the speech emotion recognizer's response by 26.13 percentage points for IndexTTS-2 and 12.97 points for CosyVoice2. The rate at which the dominant target emotion receives the largest posterior increase improves by 12.95 and 6.92 points, respectively. Human evaluation shows the same trend. The percentage of clips whose most frequent listener annotation matches the dominant target emotion increases from 54.72\% to 73.89\% for IndexTTS-2 and from 58.18\% to 77.88\% for CosyVoice2. EmoRES also obtains statistically significant naturalness preference scores of 63.80\% and 60.32\% over CoCoEmo on the two backbones. 
The component ablations in Section~\ref{sec:component-ablation} further support our hypothesis that the shared component primarily moves speech away from neutral expression, while the residual directs generation toward the requested emotion. Finally, our contributions are as follows:
\begin{itemize}
    \item To our knowledge, we conduct the first functional analysis of the shared component and category-specific residuals in emotion steering vectors for LM-based TTS.
    \item We introduce EmoRES, a training-free generalization of conventional steering that reweights the shared and residual components without learning new subspaces or updating parameters of TTS models. 
    Experiments across multiple TTS backbones, held-out evaluation, and human judgments consistently demonstrate improved emotional control.
\end{itemize}

%% file: related_work.tex
\noindent\textbf{Emotion-controllable text-to-speech.}
Emotion-controllable TTS commonly uses categorical labels, continuous emotion embeddings, natural-language instructions, or expressive reference speech. Label-based approaches such as EmoSphere++~\citep{cho2025emosphere} represent emotion type and intensity in a continuous affective space, while PromptTTS~\citep{guo2023prompttts} and EmoVoice~\citep{yang2025emovoice} use textual descriptions to specify speaking style. Recent zero-shot systems provide similar controls through explicit emotion embeddings, as in IndexTTS-2~\citep{indextts2}, or natural-language instructions, as in CosyVoice2~\citep{cosyvoice2}. Mixed-emotion synthesis~\citep{zhou2022speech} has also been studied using models trained specifically for compound emotional expression. These approaches provide useful control interfaces, but generally depend on emotion-aware training, specialized conditioning modules, or expressive reference signals.

\noindent\textbf{Vector steering for speech generation.}
Vector steering modifies intermediate representations at inference time without updating model parameters. Early work in language models constructs directions from contrasting examples and adds them to internal activations to control high-level behavior~\citep{turner2023steering,zou2023representation, rimsky2024steering}. Related approaches manipulate parameter, speaker, or style embeddings to transfer emotional characteristics in TTS~\citep{chen2024emoknob,de2026task}. EmoSteer-TTS applies difference-of-means vectors within flow-matching TTS models and supports emotion conversion, intensity control, erasure, and vector composition~\citep{xie2025emosteer}. CoCoEmo instead identifies the speech language model as an effective steering site and combines categorical directions to produce quantitative mixed-emotion speech~\citep{cocoemo}. Subsequent geometric~\citep{wang2026geometric} analysis finds that speech language model representations provide more separable and speaker-invariant emotion directions than flow-matching representations. Our work adopts this established extraction and inference-time vector steering framework of CoCoEmo~\citep{cocoemo}, but instead examines how categorical directions of emotions are composed and separates the shared displacement from the request-dependent residual.

\noindent\textbf{Composition and representation structure.}
Existing TTS steering methods construct compound emotions by taking weighted sums of categorical directions~\citep{xie2025emosteer,cocoemo}. This approach treats each direction as an indivisible unit and does not account for structure shared across the emotional vectors. Related work on multi-attribute language-model steering has addressed interference among directions through orthogonal constraints and learned shared and attribute-specific subspaces~\citep{jiang2026msrs}. Such methods require learning or optimizing new steering subspaces and do not examine the common displacement present in mean-difference emotion vectors. Our proposed EmoRES method instead uses an exact decomposition around the centroid of the emotion-specific mean activations. Under convex composition, the shared component retains a fixed coefficient, while only the residual changes with the requested proportions. EmoRES exposes separate controls for these two components without learning a new subspace or modifying the weights of the TTS backbone.

%% file: method.tex
In this work, emotional speech is generated by intervening on the internal activations of a frozen text-to-speech backbone. The procedure is composed of two parts, the emotion vector extraction phase and the vector steering phase. 
The remainder of this section specifies the two stages, and then examines the structure of the injected vectors, which admits a decomposition that CoCoEmo~\citep{cocoemo} leaves unexploited and that motivates the proposed method.


\noindent\textbf{Steering vector extraction.}
In this phase, a set of emotion vectors is extracted offline from expressive emotional corpora by contrasting the model activations of emotional and neutral speech recordings. This is a one-time computation and these vectors are then reused across all utterances for generation during inference. Let $\bar h_{e}$ denote that activation averaged over all recordings for each non-neutral emotion $e$, and $\bar h_{\mathrm{neu}}$ the mean of the neutral reference activations.
The vector for an emotion is the difference of the two means,
\begin{equation}
v_{e} \;=\; \bar h_{e} - \bar h_{\mathrm{neu}} ,
\label{eq:vemo}
\end{equation}
so that $v_{e}$ is the average displacement in activation space that separates speech carrying emotion from neutral speech. Extraction requires no gradient computation and no model weight updates.

\noindent\textbf{Steering.}
During inference, the emotion vector is added back into the activation at the same site it was extracted. The activation $h$ is displaced along the requested steering direction:
\begin{equation}
h \;\leftarrow\; h + \alpha v ,
\label{eq:add}
\end{equation}
where $v$ is the steering vector for the requested emotion and $\alpha$ a scalar for controlling the steering strength. 
Applied directly, Eq.~(\ref{eq:add}) alters the magnitude of the hidden state as well as its direction, and large steering magnitudes may move activations outside the distribution encountered during training. The magnitude of each position is therefore restored after the addition, and the update actually applied is
\begin{equation}
h \;\leftarrow\; \lVert h\rVert \cdot \frac{h+\alpha v}{\lVert h+\alpha v\rVert} ,
\label{eq:inject}
\end{equation}
where $\lVert\cdot\rVert$ denotes the Euclidean norm over the hidden dimension.
The direction of the hidden state carries the emotional content, while its norm is left untouched. The backbone parameters remain frozen throughout, and generation requires no target-emotion reference recording.

\noindent\textbf{Shared and residual components.}
The central observation of this work is that the emotion vector
$v_{e} = \bar h_{e} - \bar h_{\mathrm{neu}}$ can be decomposed into two functionally distinct components. 
Let $\mathcal{E}$ denote the set of non-neutral emotion categories. We define
  $
  \bar h_c=\frac{1}{|\mathcal{E}|}
  \sum_{e\in\mathcal{E}}\bar h_{e}
  $
as the unweighted centroid of the mean activations of emotional speech.
Inserting $\bar h_c$ into Eq.~(\ref{eq:vemo}) splits the vector into two parts,
\begin{equation}
v_{e} =\bar h_{e} - \bar h_{\mathrm{neu}}
\;=\;
\underbrace{\bar h_c - \bar h_{\mathrm{neu}}}_{\text{shared}}
\;+\;
\underbrace{\bar h_{e} - \bar h_c}_{\text{residual}} .
\label{eq:split}
\end{equation}

The two parts are interpreted differently. The shared component $\bar h_c - \bar h_{\mathrm{neu}}$ is the same vector for every request, and it carries the activation away from neutral speech and toward the centroid of emotional activations. It determines whether the speech sounds emotional, but not which emotion it
conveys. The categorical residual component $\bar h_{e} - \bar h_c$ is the part that depends on the requested emotion, and therefore carries the information distinguishing one emotion from another.

\noindent\textbf{EmoRES (Emotion residual-enhanced steering).}
In the setting of CoCoEmo~\citep{cocoemo}, steering with $v_{e}$ couples the shared shift and categorical residual. The single steering strength $\alpha$ scales both components by the same factor, fixing their relative weighting. We propose to break this coupling by assigning an independent coefficient to each component, resulting in the residual-enhanced steering vector $v_{\mathrm{RES}}$:
\begin{equation}
v_{\mathrm{RES}}
\;=\;
\lambda_c\big(\bar h_c - \bar h_{\mathrm{neu}}\big)
\;+\;
\lambda_r\big(\bar h_{e} - \bar h_c\big),
\label{eq:res}
\end{equation}
where $\lambda_c$ and $\lambda_r$ control the shared and residual components, respectively. This formulation allows the neutral-to-centroid shift and the category-specific contrast to be adjusted independently.
We denote this rule by $\mathrm{RES}(\lambda_c,\lambda_r)$ and refer to it as residual-enhanced steering.
At $\lambda_c=\lambda_r=1$ the centroid cancels and
Eq.~(\ref{eq:res}) reduces exactly to $v_{e}$, so CoCoEmo's conventional steering is a special case of the proposed family rather than a distinct rule. 

\noindent\textbf{Mixture targets.}
Human speech can express multiple emotions simultaneously, so the steering rule should support mixed-emotion targets. Let $p_{e}\geq 0$ denote the requested proportion of emotion $e\in\mathcal E$, with $\sum_{e\in\mathcal E}p_{e}=1$. Using these proportions, the activation mixtures and steering vectors are
  \begin{equation}
  \bar h_p
  =
  \sum_{e\in\mathcal E}
  p_{e}\bar h_{e},
  \qquad
  v_{\mathrm{mixRES}}
  =
  \lambda_c\big(\bar h_c-\bar h_{\mathrm{neu}}\big)
  +
  \lambda_r\big(\bar h_p-\bar h_c\big).
  \label{eq:mix}
  \end{equation}
  At inference time, $v_{\mathrm{mixRES}}$ is used as the steering vector $v$ in Eq.~(\ref{eq:add}).
  The shared shift remains constant across requested emotions, while the requested proportions determine the emotional residual. See Appendix~\ref{sec:app-geometry} for a geometric interpretation of EmoRES.

%% file: exp_setup.tex
\subsection{Datasets}
\noindent\textbf{Steering vector extraction.}
Following the procedure of CoCoEmo~\citep{cocoemo}, all steering directions are extracted from a fixed library of neutral--emotional utterance pairs drawn from three corpora: ESD~\citep{esd}, CREMA-D~\citep{cremad} and RAVDESS~\citep{ravdess}. Within a pair the two utterances share both speaker and lexical content, so that the difference between them isolates emotional variation from speaker identity and text content. Both sides of every pair are screened by a speech-emotion-recognition gate and a signal-quality gate, and a pair is discarded whole if either side fails. Partial pairs are never used, since a difference taken between two different sets of utterances would not be a paired contrast. The library spans the four emotions evaluated in this work, namely \emph{angry}, \emph{happy}, \emph{sad} and \emph{surprise}, with \emph{neutral} as the reference pole. More experimental details on steering vector extraction can be found in Appendix~\ref{sec:app-exp-detail}.

\noindent\textbf{Evaluation sets.}
We employ a corpus-level development--test split for evaluation. CREMA-D~\citep{cremad} serves as the development set, annotated with per-utterance emotion mixtures over the emotions \emph{angry}, \emph{happy}, and \emph{sad}. All hyperparameter tuning and design choices, including steering coefficients, are conducted exclusively on this corpus. To evaluate out-of-distribution transferability, IEMOCAP~\citep{iemocap} is strictly held out as the test set, contributing no speakers or recordings to steering vector construction. Annotated with per-utterance mixtures over the emotions \emph{angry}, \emph{happy}, \emph{surprise}, and \emph{sad}, IEMOCAP assesses whether parameters optimized on the development set transfer effectively across domains.

\subsection{Model steering}
We evaluate two frozen zero-shot TTS backbones chosen to differ in how emotion is conditioned during inference. IndexTTS-2~\citep{indextts2} is an embedding-conditioned model that takes an explicit emotion embedding as a conditioning input. For IndexTTS-2, we steer layers $1$, $6$ and $8$ of its semantic language model. CosyVoice~2~\mbox{\citep{cosyvoice2}} is an instruction-based model that conditions on a natural-language style prompt. For CosyVoice~2, we steer layers $14$ and $17$ of its text-to-token language model. Both layer sets are the top-$K$ most emotion-separable layers published by CoCoEmo~\citep{cocoemo} for these backbones. In both cases, steering is applied at the attention output, which is also the site at which the steering vectors are extracted. No weights are updated and each model's native emotion conditioning is left at its default, so any measured effect is attributable to the steering vector alone. 

In the experiments of this work, to separate improvements from overall steering magnitude, the vector produced by each method before steering is rescaled to the mean Euclidean norm of the original emotion vectors in Eq. (\ref{eq:vemo}) at the corresponding layer. Consequently, at a fixed steering strength $\alpha$, all steered conditions inject vectors with the same norm and differ only in direction.

\subsection{Evaluation Metrics}
\label{ssec:metrics}

\subsubsection{Objective evaluation}
We assess emotional expression using target emotion probability (\textbf{TEP}) and weighted anchored emotion similarity (\textbf{E-SIM}), speaker preservation using speaker similarity (\textbf{S-SIM}), and intelligibility using word error rate (\textbf{WER}). For mixed-emotion requests, we additionally report Spearman rank correlation ($\bm{\rho}$) and hit rate (\textbf{H-Rate}) to measure whether changes in the speech emotion recognizer posteriors follow the requested ordering and dominant emotion. Full definitions and evaluation details are provided in Appendix~\ref{app:obj_metric}.

\subsubsection{Subjective evaluation}
We conduct two human evaluations on IEMOCAP, emotion labeling and naturalness preference. For emotion labeling, annotators listen to one clip at a time and assign the emotion they perceive. The labels collected for each clip form an empirical emotion distribution, from which \textbf{Dom-hit} and \textbf{Fidelity} are computed. For naturalness, annotators listen to outputs from CoCoEmo and our proposed EmoRES method for the same sentence in randomized order and choose whether EmoRES is more natural, CoCoEmo is more natural, or the two sound about the same.
For each subjective evaluation task, each sample received at least 9 annotations from different annotators.
We briefly describe each metric below, with complete definitions and evaluation details provided in Appendix~\ref{app:sub_metric}.

\noindent\textbf{Dom-hit (Dominant emotion hit rate):} The fraction of clips whose most frequent annotated emotion label is the utterance's dominant target emotion. 
\noindent\textbf{Fidelity:} The agreement between the distribution of the annotated emotion labels
over a clip and the target emotion mixture.
Fidelity is sensitive to the whole mixture and therefore penalizes a generated clip that reaches the dominant emotion while suppressing the rest of the requested emotions.
\noindent\textbf{Naturalness preference:} The tie-adjusted fraction of blind pairwise comparisons in which listeners judged the proposed system more natural than the CoCoEmo baseline. It is an ordinal comparison of two systems rather than an absolute quality score, and it detects a loss of naturalness paid for emotional control.

\subsection{Baselines}
We compare against three baselines, all decoded with the same manifests, prompts and frozen backbones. \emph{No-steer} is the unmodified TTS backbone for emotional speech generation without steering. By default, generation uses each backbone's native emotion-conditioning interface, namely emotion embeddings for IndexTTS-2 and natural-language instructions for CosyVoice2, unless otherwise specified. This baseline fixes the intelligibility and speaker-identity operating point, and is the reference against which metrics $\rho$ and H-Rate are computed. 
\emph{CoCoEmo}~\citep{cocoemo} serves as the emotion vector steering baseline and is recovered exactly at $\lambda_c=\lambda_r=1$.
\emph{Random steer} injects an isotropic Gaussian direction at the same layers and site, rescaled to the norm of the vector it replaces, so only the direction changes. It isolates the contribution of the extracted direction from that of the injected magnitude alone, and thus establishes the lower bound that any direction-specific claim must exceed. Results for random steering are provided in Appendix~\ref{sec:app-ablmix} due to page limitations.

%% file: results.tex
We evaluate EmoRES through objective and subjective comparisons in Section~\ref{sec:main-results}, component ablations in Section~\ref{sec:component-ablation}, and visualization of speech emotion embeddings in Section~\ref{subsec:visual}.
Additional experiments examine random-direction controls in Appendix~\ref{sec:app-ablmix}, single-emotion requests for emotional speech prompts in Appendix~\ref{sec:app-single-emo-prompt}, steering without native emotion conditioning in Appendix~\ref{sec:app-no-conditioning}, and evaluating with different speech emotion recognizers in Appendix~\ref{sec:app-3-ser}.
\input{figures/grid6_vs_alpha_abnray}

\subsection{Residual-Enhanced Vectors}
\label{sec:main-results}
Figure~\ref{fig:lrray} plots each metric across steering strengths $\alpha\in\{1,\ldots,6\}$ for various residual strengths $\lambda_r\in\{0,\ldots,5\}$ with IndexTTS-2, evaluated on the IEMOCAP evaluation set. 
With the residual strength set to $\lambda_r=2$, shown in green, increasing $\alpha$ monotonically improves all four emotion metrics. Compared with CoCoEmo ($\lambda_r=1$), it achieves higher TEP, $\rho$, H-Rate, E-SIM, and S-SIM across all evaluated steering strengths while generally yielding lower WER.
Increasing the residual strength to $\lambda_r=3$, shown by the red curve, outperforms CoCoEmo ($\lambda_r=1$) on every reported metric across all evaluated steering strengths while consistently yielding lower WER than CoCoEmo.
In contrast, shared-only steering ($\lambda_r=0$) generally deteriorates across all metrics as $\alpha$ increases. CoCoEmo largely saturates beyond $\alpha=3$ on emotion metrics, while S-SIM decreases and WER increases. Increasing $\lambda_r$ beyond 3 provides smaller additional gains, suggesting diminishing returns.

\input{tables/main_results}

Table~\ref{tab:mixed} shows the emotional speech generation results for IndexTTS-2 and CosyVoice2 on the in-distribution CREMA-D development set and the out-of-distribution IEMOCAP set, comparing EmoRES with CoCoEmo and no steering. All steered rows share $\lambda_c = 1$, with the steering vectors normalized to the average length of $v_{e}$, and vary only in the residual steering strength $\lambda_r$. Consequently, performance differences cannot be attributed to unequal steering magnitudes.

The no-steer baseline uses each original model's native conditioning, with emotion embeddings for IndexTTS-2 and emotion instructions for CosyVoice2, but without vector steering. Despite this conditioning, its lower TEP and E-SIM relative to the steered systems indicate that native conditioning alone provides weaker alignment with the requested emotions.
Compared with CoCoEmo, EmoRES yields its largest improvements in $\rho$ and H-Rate. On CREMA-D, $\rho$ increases from 25.24\% to 45.87\% for IndexTTS-2 and from 32.48\% to 51.94\% for CosyVoice2, while H-Rate increases from 76.00\% to 82.49\% and from 78.92\% to 84.76\%, respectively. On IEMOCAP, $\rho$ increases from 22.00\% to 48.13\% for IndexTTS-2 and from 39.13\% to 52.10\% for CosyVoice2, while H-Rate increases from 64.34\% to 77.29\% and from 70.90\% to 77.82\%, respectively. These results show that changes in the speech emotion recognizer's posterior more closely follow the requested rank ordering and more often assign the largest increase to the dominant target emotion. TEP and E-SIM also improve in every comparison, indicating higher average posterior mass on the requested emotion set and stronger similarity to the target-weighted emotion anchors. EmoRES further improves S-SIM in every comparison and reduces WER under most settings.

Table~\ref{tab:subj}(a) reports the subjective emotion labeling results on IEMOCAP. Without vector steering, the native controls provide weak agreement with the requested emotions, particularly for IndexTTS-2, which obtains a Dom-hit of 17.78\% and a Fidelity of 25.04\%. Compared with CoCoEmo, EmoRES raises Dom-hit from 54.72\% to 73.89\% for IndexTTS-2 and from 58.18\% to 77.88\% for CosyVoice2. Thus, the emotion most frequently identified by listeners matches the dominant target emotion more often under EmoRES. Fidelity increases from 60.05\% to 66.32\% for IndexTTS-2 and from 50.80\% to 59.60\% for CosyVoice2. 
The improvements in both Dom-hit and Fidelity are statistically significant for both backbones, with paired 95\% confidence intervals for the differences between EmoRES and CoCoEmo excluding zero.
The simultaneous improvement in Fidelity shows that the gain in Dom-hit is accompanied by better agreement with the complete target mixture, rather than only stronger expression of its dominant emotion. 
These subjective trends align with Table~\ref{tab:mixed}, where EmoRES also improves H-Rate and $\rho$ over CoCoEmo for both models on IEMOCAP.

Table~\ref{tab:subj}(b) reports the naturalness comparison between CoCoEmo and EmoRES. EmoRES obtains preference scores of 63.80\% for IndexTTS-2 and 60.32\% for CosyVoice2. 
Both scores are significantly above the 50\% indifference point, with 95\% confidence intervals of $[61.70, 65.91]$ and $[58.08, 62.56]$, respectively.
Overall, the subjective evaluation results indicate that, relative to CoCoEmo, EmoRES improves emotional control while better preserving perceived naturalness.

\input{tables/subjective}

\subsection{Shared and Residual Components}
\label{sec:component-ablation}


\input{tables/ablmix}

Table~\ref{tab:ablmix_} isolates the roles of the shared and residual components by holding $\alpha$ and the steering-vector norm fixed while varying only $\lambda_c$ and $\lambda_r$. Shared-only steering improves E-SIM and TEP over no steering, but produces the lowest $\rho$ and H-Rate among the steered conditions for both models. In particular, $\rho$ falls to 5.30\% for IndexTTS-2 and 7.20\% for CosyVoice2. 
As a separate diagnostic, we find that shared-only steering reduces the speech emotion recognizer's neutral posterior from 36.51\% to 7.49\% for IndexTTS-2.
Together, these results show that the shared component moves the output away from neutral, but provides imprecise information about the requested emotion mixture.

Residual-only steering exhibits the opposite pattern. Relative to the conventional setting of CoCoEmo at $\lambda_c=\lambda_r=1$, it improves $\rho$ and H-Rate but reduces E-SIM and TEP for both models. Thus, neither component is sufficient alone. The shared component increases posterior mass on the requested emotion set and similarity to the emotion anchors, but carries little information about the requested mixture. The residual provides category-specific contrast, but when used alone yields lower TEP and E-SIM than the combined direction.
Moreover, simply including both components with equal weights remains suboptimal. Retaining the shared component while increasing the relative residual strength to $\lambda_r=3$ produces the best emotion-control results in the table. Since all conditions are norm-matched, this improvement arises from rebalancing the two components rather than increasing the overall steering magnitude.

\subsection{Visualization of speech emotion embeddings}
\label{subsec:visual}
\input{figures/res_tsne}

The t-SNE plot~\citep{tsne} in Figure~\ref{fig:tsne} illustrates how vector steering changes the distribution of generated utterances in the emotion2vec~\citep{emotion2vec} embedding space. To quantify emotion-specific organization, we report two complementary measures computed from the emotion2vec embeddings. Probe accuracy is the balanced accuracy of a logistic regression classifier trained to predict the requested emotion, and therefore measures how emotion identity can be recovered from the embeddings. The cosine silhouette score measures within-emotion cohesion relative to separation from other emotions. Values near one indicate compact, well-separated groups, values near zero indicate overlapping groups, and negative values indicate that samples are often closer to another emotion than to their own.

The no-steer embeddings in panel (b) exhibit substantial overlap among the requested emotions. Although native conditioning makes emotion identity partially recoverable, as reflected by a probe accuracy of 56.9\%, the slightly negative silhouette score indicates that it does not form compact emotion-specific clusters. Shared-only steering in panel (c) produces similar results, with 57.6\% probe accuracy and a silhouette score near zero.
In contrast, residual-only steering in panel (d) forms substantially more distinct clusters, achieving 81.4\% probe accuracy and a silhouette score of 0.379. This supports the interpretation that the residual component carries the category-specific contrasts between emotions.

CoCoEmo in panel (e) combines both components but retains some overlap between the emotion clusters. Our EmoRES method in panel (f) strengthens the residual while retaining the shared component, producing the highest probe accuracy of 86.9\% and silhouette score of 0.460. 
Compared with CoCoEmo in panel (e), EmoRES in panel (f) produces more compact and distinct emotion-dependent clusters. In particular, the sad and happy clusters are more clearly separated under EmoRES.
This visual pattern is consistent with the higher probe accuracy and cosine silhouette score achieved by EmoRES, indicating that the requested emotions are more distinguishable and that utterances sharing the same target are grouped more closely.
These results suggest that strengthening the residual component improves category-specific emotion structure in embedding space.


%% file: figures/grid6_vs_alpha_abnray.tex
\begin{figure}[t]
\centering
\includegraphics[width=16cm]{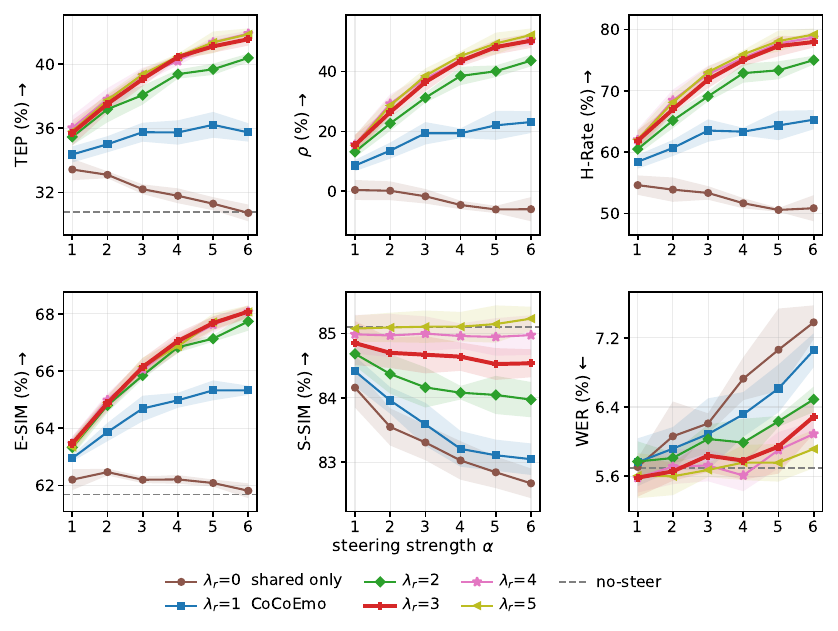}
\caption{Residual weight against steering strength $\alpha$ on IEMOCAP (OOD) with IndexTTS-2 for each objective metric. Each curve is fixed at $\lambda_c = 1$ with $\lambda_r$ ablated. Every arm is norm-matched, so at a given $\alpha$, all arms inject the same steering vector norm and differ only in the direction. $\lambda_r = 1$ is CoCoEmo's steering method. Shaded bands are $\pm$ one standard deviation over five different runs. 
}
\label{fig:lrray}
\end{figure}

%% file: tables/main_results.tex
\begin{table*}[t]
\centering
\small
\setlength{\tabcolsep}{4pt}
\caption{Mixed-emotion evaluation on CREMA-D (dev) and IEMOCAP (OOD). All steered rows use $\alpha=5$ and $\lambda_c=1$.
CoCoEmo is the $\lambda_r=1$ special case. Values are percentages, reported as
mean $\pm$ standard deviation over five replicates. Best results
among the synthesis rows are \textbf{bold}, second-best are \underline{underlined}.}
\label{tab:mixed}
\begin{tabular}{lcccccc}
\toprule
Method & E-SIM$\uparrow$ & TEP$\uparrow$ & $\rho\uparrow$ & H-Rate$\uparrow$ & S-SIM$\uparrow$ & WER$\downarrow$ \\
\midrule
\multicolumn{7}{c}{\textit{In-distribution evaluation on CREMA-D (dev)}} \\
\midrule
Ground Truth & 72.56 & 55.51 & -- & -- & 87.86 & 1.08 \\
\midrule
\multicolumn{7}{l}{\textbf{IndexTTS-2}} \\
No-steer & 55.00$\pm$0.36 & 16.83$\pm$1.39 & -- & -- & \textbf{87.18$\pm$0.16} & \textbf{6.30$\pm$0.05} \\
CoCoEmo & \underline{64.46$\pm$0.40} & \underline{38.69$\pm$1.21} & \underline{25.24$\pm$4.52} & \underline{76.00$\pm$1.78} & 85.77$\pm$0.22 & 7.60$\pm$0.14 \\
EmoRES, $\lambda_r=3$ & \textbf{64.80$\pm$0.25} & \textbf{39.98$\pm$0.88} & \textbf{45.87$\pm$6.99} & \textbf{82.49$\pm$2.34} & \underline{86.90$\pm$0.21} & \underline{6.78$\pm$0.16} \\
\midrule
\multicolumn{7}{l}{\textbf{CosyVoice2}} \\
No-steer & 59.70$\pm$0.36 & 34.01$\pm$0.96 & -- & -- & \textbf{87.36$\pm$0.21} & \textbf{0.53$\pm$0.06} \\
CoCoEmo & \underline{74.16$\pm$0.09} & \underline{67.43$\pm$0.40} & \underline{32.48$\pm$2.17} & \underline{78.92$\pm$0.54} & 86.30$\pm$0.12 & \underline{1.21$\pm$0.17} \\
EmoRES, $\lambda_r=3$ & \textbf{74.71$\pm$0.24} & \textbf{68.80$\pm$0.56} & \textbf{51.94$\pm$6.30} & \textbf{84.76$\pm$2.07} & \underline{86.73$\pm$0.18} & 1.30$\pm$0.10 \\
\midrule
\multicolumn{7}{c}{\textit{Out-of-distribution evaluation on IEMOCAP (OOD)}} \\
\midrule
Ground Truth & 60.68 & 39.82 & -- & -- & 87.06 & 11.95 \\
\midrule
\multicolumn{7}{l}{\textbf{IndexTTS-2}} \\
No-steer & 61.69$\pm$0.23 & 30.76$\pm$0.44 & -- & -- & \textbf{85.09$\pm$0.16} & \textbf{5.69$\pm$0.20} \\
CoCoEmo & \underline{65.32$\pm$0.35} & \underline{36.22$\pm$0.80} & \underline{22.00$\pm$4.83} & \underline{64.34$\pm$2.39} & 83.11$\pm$0.24 & 6.61$\pm$0.28 \\
EmoRES, $\lambda_r=3$ & \textbf{67.67$\pm$0.23} & \textbf{41.11$\pm$0.39} & \textbf{48.13$\pm$2.44} & \textbf{77.29$\pm$1.67} & \underline{84.52$\pm$0.25} & \underline{5.94$\pm$0.31} \\
\midrule
\multicolumn{7}{l}{\textbf{CosyVoice2}} \\
No-steer & 56.74$\pm$0.21 & 28.92$\pm$0.17 & -- & -- & \textbf{87.49$\pm$0.14} & \textbf{3.21$\pm$0.22} \\
CoCoEmo & \underline{63.43$\pm$0.22} & \underline{37.44$\pm$0.36} & \underline{39.13$\pm$1.56} & \underline{70.90$\pm$0.81} & 86.42$\pm$0.14 & 4.51$\pm$0.17 \\
EmoRES, $\lambda_r=3$ & \textbf{65.34$\pm$0.22} & \textbf{40.12$\pm$0.29} & \textbf{52.10$\pm$1.05} & \textbf{77.82$\pm$0.68} & \underline{86.81$\pm$0.13} & \underline{3.61$\pm$0.12} \\
\bottomrule
\end{tabular}
\end{table*}

%% file: tables/subjective.tex
\begin{table}[t]
\centering
\footnotesize
\setlength{\tabcolsep}{2pt}
\caption{Subjective evaluation on IEMOCAP, comparing CoCoEmo ($\lambda_r=1$) against EmoRES ($\lambda_r=3$).
$\Delta$ is the difference between EmoRES and CoCoEmo. $^{*}$ denotes statistical significance.}
\label{tab:subj}
\scriptsize
\begin{minipage}[t]{0.50\textwidth}
\centering
\textbf{(a)} Emotion labeling\\[2pt]
\begin{tabular}{lcccc}
\toprule
 & \multicolumn{2}{c}{IndexTTS-2} & \multicolumn{2}{c}{CosyVoice2} \\
\cmidrule(lr){2-3}\cmidrule(lr){4-5}
Method & Dom-hit $\uparrow$ & Fidelity $\uparrow$ & Dom-hit $\uparrow$ & Fidelity $\uparrow$ \\
\midrule
No-steer & 17.78\,\% & 25.04\,\% & 32.42\,\% & 33.28\,\% \\
CoCoEmo & 54.72\,\% & 60.05\,\% & 58.18\,\% & 50.80\,\% \\
EmoRES & \textbf{73.89\,\%} & \textbf{66.32\,\%} & \textbf{77.88\,\%} & \textbf{59.60\,\%} \\
\midrule
$\Delta$ & $+$19.17\,$^{*}$ & $+$6.27\,$^{*}$ & $+$19.70\,$^{*}$ & $+$8.80\,$^{*}$ \\
95\% CI of $\Delta$ & [13.06, 25.56] & [3.44, 8.94] & [13.33, 26.07] & [5.91, 11.81] \\
\bottomrule
\end{tabular}
\end{minipage}\hfill
\begin{minipage}[t]{0.49\textwidth}
\centering
\textbf{(b)} Naturalness A/B\\[2pt]
\begin{tabular}{lcc}
\toprule
 & IndexTTS-2 & CosyVoice2 \\
\midrule
EmoRES preference score $\uparrow$ & \textbf{63.80\,\%}$^{*}$ & \textbf{60.32\,\%}$^{*}$ \\
\quad 95\% CI & [61.70, 65.91] & [58.08, 62.56] \\
\quad Choose EmoRES & \textbf{50.83\,\%} & \textbf{41.78\,\%} \\
\quad About the same & 25.93\,\% & 37.07\,\% \\
\quad Choose CoCoEmo & 23.24\,\% & 21.14\,\% \\
\bottomrule
\end{tabular}
\end{minipage}
\end{table}

%% file: tables/ablmix.tex
\begin{table}[t]
\centering
\small
\setlength{\tabcolsep}{3pt}
\caption{Component ablation on CREMA-D in the mixed setting. Every steered row uses steering strength $\alpha = 5$, norm-matched to the same steering vector norm and differs from the others only in how that norm is split between the shared and residual components and in the direction each component points. Mean $\pm$ one standard deviation over five speaker-prompt runs. Best per column and backbone in bold. See Appendix~\ref{sec:app-ablmix} for the design and for the row contrasts.}
\label{tab:ablmix_}
\begin{tabular}{lcccccccc}
\toprule
 Method & $\lambda_c$ & $\lambda_r$ & E-SIM $\uparrow$ & TEP $\uparrow$ & $\rho$ $\uparrow$ & H-Rate $\uparrow$ & S-SIM $\uparrow$ & WER $\downarrow$ \\
\midrule
Ground Truth & -- & -- & 72.56 & 55.51 & -- & -- & 87.86 & 1.08 \\
\midrule
\multicolumn{9}{l}{\textbf{IndexTTS-2}} \\
No-steer & -- & -- & 55.00$\pm$0.36 & 16.83$\pm$1.39 & -- & -- & 87.18$\pm$0.16 & \textbf{6.30$\pm$0.05} \\
Residual only & $0$ & $1$ & 63.05$\pm$0.29 & 35.88$\pm$0.77 & 45.16$\pm$5.92 & 82.16$\pm$2.06 & \textbf{87.64$\pm$0.18} & 6.36$\pm$0.06 \\
Shared \& Residual & $1$ & $1$ & 64.46$\pm$0.40 & 38.69$\pm$1.21 & 25.24$\pm$4.52 & 76.00$\pm$1.78 & 85.77$\pm$0.22 & 7.60$\pm$0.14 \\
Shared only & $1$ & $0$ & 59.51$\pm$0.33 & 26.64$\pm$1.18 & 5.30$\pm$7.86 & 69.62$\pm$2.21 & 85.13$\pm$0.22 & 6.69$\pm$0.26 \\
Shared \& Residual & $1$ & $3$ & \textbf{64.80$\pm$0.25} & \textbf{39.98$\pm$0.88} & \textbf{45.87$\pm$6.99} & \textbf{82.49$\pm$2.34} & 86.90$\pm$0.21 & 6.78$\pm$0.16 \\
\midrule
\multicolumn{9}{l}{\textbf{CosyVoice2}} \\
No-steer & -- & -- & 59.70$\pm$0.36 & 34.01$\pm$0.96 & -- & -- & \textbf{87.36$\pm$0.21} & \textbf{0.53$\pm$0.06} \\
Residual only & $0$ & $1$ & 72.21$\pm$0.27 & 63.82$\pm$0.56 & 39.94$\pm$5.04 & 80.54$\pm$1.53 & 87.21$\pm$0.16 & 1.14$\pm$0.15 \\
Shared \& Residual & $1$ & $1$ & 74.16$\pm$0.09 & 67.43$\pm$0.40 & 32.48$\pm$2.17 & 78.92$\pm$0.54 & 86.30$\pm$0.12 & 1.21$\pm$0.17 \\
Shared only & $1$ & $0$ & 65.92$\pm$0.26 & 47.53$\pm$0.68 & 7.20$\pm$5.03 & 70.16$\pm$2.14 & 85.77$\pm$0.13 & 2.30$\pm$0.19 \\
Shared \& Residual & $1$ & $3$ & \textbf{74.71$\pm$0.24} & \textbf{68.80$\pm$0.56} & \textbf{51.94$\pm$6.30} & \textbf{84.76$\pm$2.07} & 86.73$\pm$0.18 & 1.30$\pm$0.10 \\
\bottomrule
\end{tabular}
\end{table}

%% file: figures/res_tsne.tex
\begin{figure}[t]
\centering
\includegraphics[width=16cm]{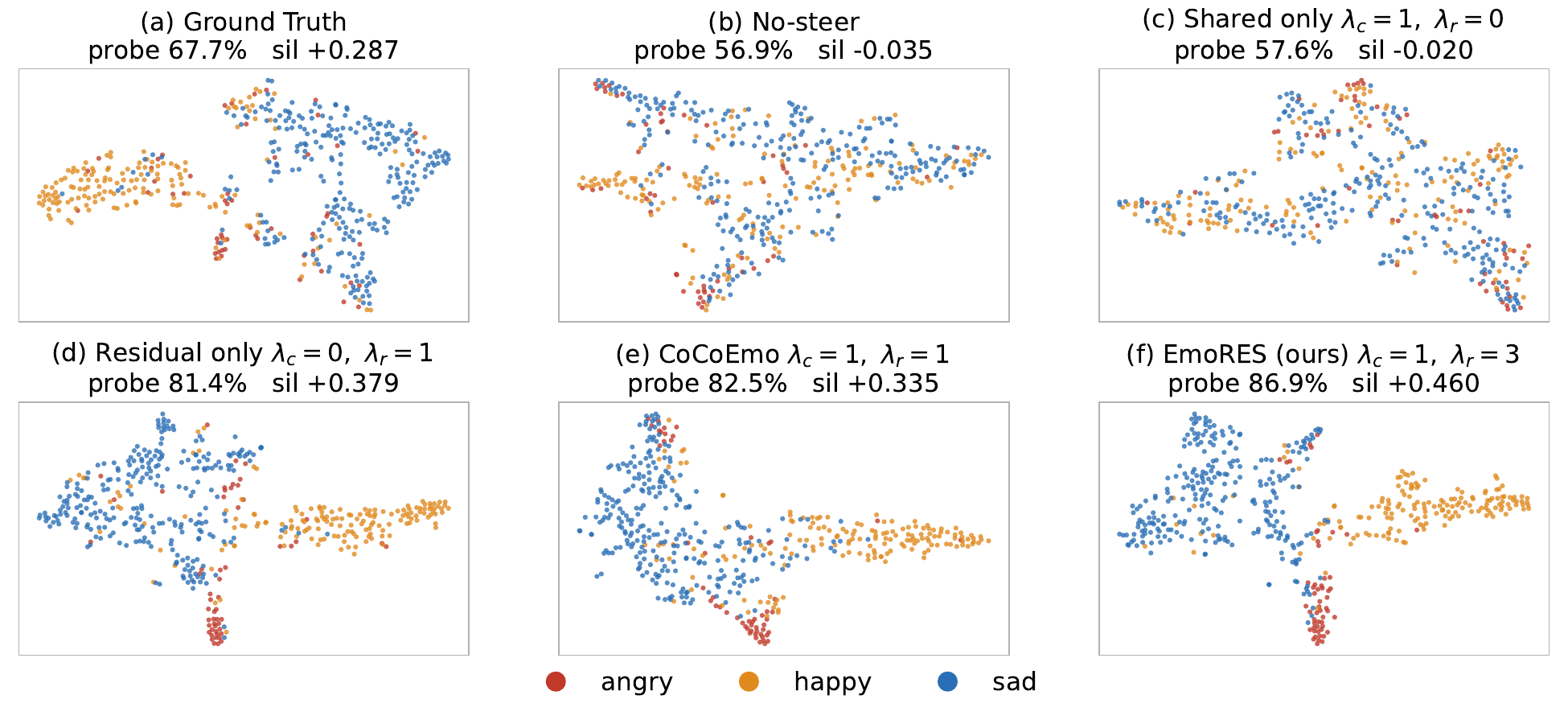}
\caption{Emotion2vec embedding t-SNE plots for utterances generated by IndexTTS-2 on the single-emotion CREMA-D subset ($n=486$). Each point is colored by its requested emotion. Values above each panel report the accuracy of probing and the mean cosine silhouette score.}
\label{fig:tsne}
\vspace{-10pt}
\end{figure}

%% file: conclusions.tex
This work shows that emotion steering vectors contain shared and residual components with distinct functions. EmoRES controls these components independently and strengthens the residual relative to the shared component without fine-tuning. Norm-matched experiments on IndexTTS-2 and CosyVoice2 demonstrate improved emotion control on both CREMA-D and IEMOCAP, with component ablations and human evaluations supporting the proposed interpretation. Future work will extend EmoRES to token-level or segment-level control for emotions that vary within an utterance and test whether the shared-residual decomposition generalizes across additional emotions, languages, speech corpora, and TTS architectures.

%% file: appendix/EmoRES_illustration.tex
\input{figures/EmoRES_illustration}


Figure~\ref{fig:emores_geometry} provides a geometric interpretation of the shared and residual components introduced in Section~\ref{section:method}. The illustration fixes $\lambda_c=1$, so the shared component first moves from the neutral mean $\bar h_{\mathrm{neu}}$ to the centroid of the emotional means
$\bar h_c$. Each categorical residual
\[
r_{e}
=
\bar h_{e}-\bar h_c
\]
then points from this centroid toward an emotion-specific mean.

Consider a mixture of angry and surprise:
\begin{equation}
\bar h_p
=
p_e\bar h_{\mathrm{angry}}
+
(1-p_e)\bar h_{\mathrm{surprise}},
\qquad
p_e\in[0,1].
\label{eq:app-two-emotion-mixture}
\end{equation}
Its residual can be written as
\begin{equation}
\bar h_p-\bar h_c
=
p_e\big(\bar h_{\mathrm{angry}}-\bar h_c\big)
+
(1-p_e)\big(\bar h_{\mathrm{surprise}}-\bar h_c\big).
\label{eq:app-mixture-residual}
\end{equation}
The mixture residual is therefore a convex combination of the two categorical residuals. As $p_e$ varies, its endpoint traces the line segment between the
angry and surprise endpoints shown in Figure~\ref{fig:emores_geometry}.

For a fixed residual strength $\lambda_r$, the scaled mixture endpoint is
\begin{align}
\bar h_c+\lambda_r(\bar h_p-\bar h_c)
&=
p_e\left[\bar h_c+\lambda_r
\big(\bar h_{\mathrm{angry}}-\bar h_c\big)\right]
\nonumber\\
&\quad+
(1-p_e)\left[\bar h_c+\lambda_r
\big(\bar h_{\mathrm{surprise}}-\bar h_c\big)\right].
\label{eq:app-scaled-mixture}
\end{align}
Consequently, every value of $\lambda_r$ defines a line segment containing all mixtures between the scaled emotion-specific endpoints. At $\lambda_r=1$,
this segment connects the original emotion means, and the complete direction reduces to
\[
\big(\bar h_c-\bar h_{\mathrm{neu}}\big)
+
\big(\bar h_p-\bar h_c\big)
=
\bar h_p-\bar h_{\mathrm{neu}},
\]
which recovers CoCoEmo. Increasing $\lambda_r$ expands the mixture segment about $\bar h_c$ and increases the relative contribution of the emotion-specific residual. In this work, EmoRES uses $\lambda_r=3$ in our main experiments.

The figure depicts the unnormalized construction in activation space. Before injection, the complete steering vector is rescaled to the common reference norm described in Section~\ref{section:exp_setup}. This rescaling preserves each vector's direction, so the experimental differences arise from
rebalancing the shared and residual components rather than merely increasing the injected norm.

%% file: figures/EmoRES_illustration.tex
\begin{figure*}[h]
  \centering
  \includegraphics[width=\textwidth]{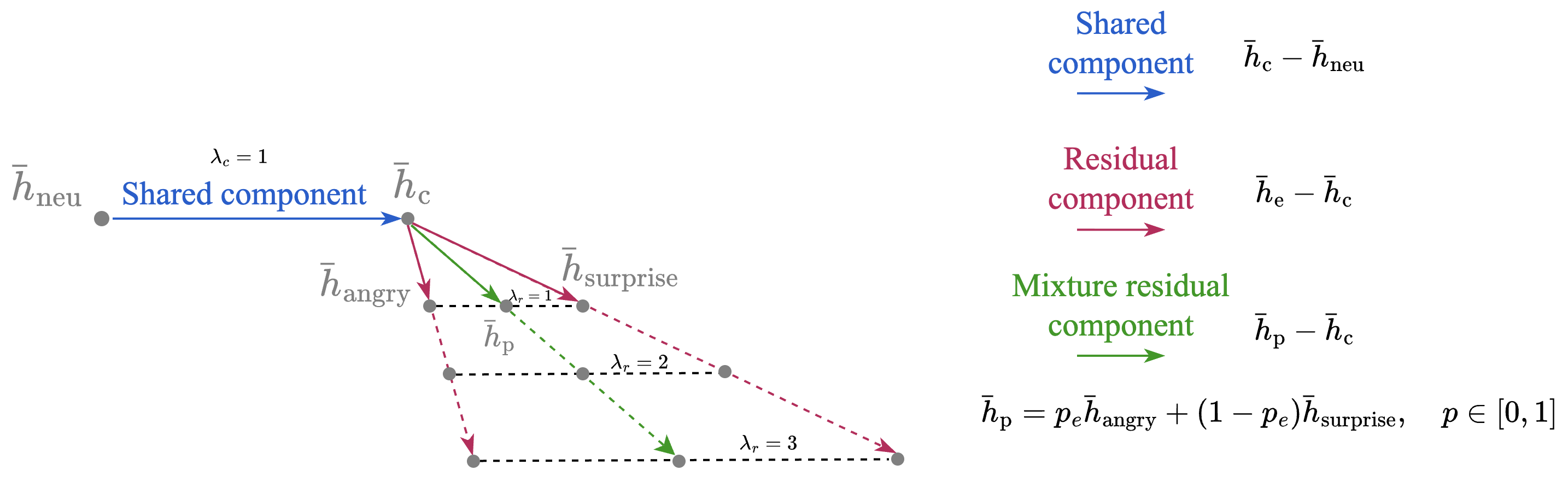}
  \caption{Geometric illustration of EmoRES in a two-dimensional projection with an example of two emotions.}
  \label{fig:emores_geometry}
\end{figure*}

%% file: appendix/metrics.tex
\subsection{Objective metrics}
\label{app:obj_metric}
All six metrics are defined per utterance and reported as the average over utterances, so the utterance index is suppressed below. Let $\mathcal{E}$ be the emotion set of the evaluation corpus, with $|\mathcal{E}|=3$ for CREMA-D and $|\mathcal{E}|=4$ for IEMOCAP. An utterance carries a target
  mixture $p_{e}\ge 0$ with $\sum_{e\in\mathcal{E}}p_{e}=1$, taken from
  the corpus rater proportions, and a target set
  $\mathcal{T}=\{e: p_{e}>0\}$. We write $P(e)$ for the posterior that
  emotion2vec+ large assigns to $e$ on the generated utterance, $P_0(e)$ for the
  same quantity on the unsteered baseline generated from the identical text and speaker prompt, and $z$
  for the unit-norm emotion2vec+ large~\citep{emotion2vec} embedding of the generated utterance.

\noindent\textbf{TEP (Target emotion probability).}
TEP is the emotion posterior that a speech-emotion recognizer assigns to an utterance's target emotions,
\begin{equation}
\mathrm{TEP}=\frac{1}{|\mathcal{T}|}\sum_{e\in\mathcal{T}}P(e) ,
\end{equation}
averaged over those targets. It measures the average posterior probability that the recognizer assigns to the requested emotions. In this work, we adopt emotion2vec+ large\footnote{\href{https://huggingface.co/emotion2vec/emotion2vec_plus_large}{emotion2vec/emotion2vec\_plus\_large}}~\citep{emotion2vec} as the speech emotion recognizer to extract probabilities for each emotion.

\noindent\textbf{E-SIM (Weighted anchored emotion similarity).}
E-SIM measures how emotionally similar the generated speech is to real recordings in embedding space.
In this work, to calculate E-SIM, we follow the procedure of EmoSteer-TTS~\citep{xie2025emosteer} by curating a set of emotional speech clips for each emotion serving as anchors.
Let $\mathcal{A}(e)$ be the anchor bank for $e$, consisting of $100$ emotion2vec+
large embeddings of ESD recordings~\citep{esd}. 
Anchor embeddings $a$ and the utterance embedding $z$
are both normalized to unit length, so their inner product $\langle z,\,a\rangle$ is the cosine
similarity between them. The similarity score for an emotion is the mean cosine between the generated utterance and that emotion's anchors, and E-SIM is the mixture-weighted sum of these scores,
\begin{equation}
S(e)=\frac{1}{|\mathcal{A}(e)|}\sum_{a\in\mathcal{A}(e)}\langle z,\,a\rangle ,
\qquad
\mathrm{E\text{-}SIM}=\sum_{e\in\mathcal{E}} p_{e}\, S(e) .
\end{equation}

\noindent\textbf{Rank correlation $\rho$.}
  Let $\Delta(e)=P(e)-P_0(e)$ be the increase in an emotion's posterior
  caused by steering. For an utterance with at least two target emotions, its targets can be ranked two
  ways: by how strongly each was requested, $p_{e}$, and by how much each actually rose,
  $\Delta(e)$. The metric is the Spearman rank correlation between these two orderings,
  \begin{equation}
  \rho=\mathrm{Spearman}\Big[\ \big\{\,p_{e},\ \Delta(e)\,\big\}_{e\in\mathcal{T}}\ \Big].
  \end{equation}
  An utterance requesting sad more strongly than angry scores $\rho=1$ when sad's posterior rises more
  than angry's, and $\rho=-1$ when angry's rises more.

  \noindent\textbf{H-Rate (Hit rate).}
H-Rate is the coarser companion to $\rho$: instead of the full ordering, it compares only the top of the
two rankings. An utterance's dominant emotion is the one requested most strongly, and the utterance
counts as a hit when that same emotion also shows the largest increase,
\begin{equation}
\mathrm{hit}=\mathbf{1}\Big[\ \arg\max_{e\in\mathcal{T}}\Delta(e)\ =\
\arg\max_{e\in\mathcal{T}}p_{e}\ \Big].
\end{equation}
H-Rate is the fraction of hits over the utterances on which it is defined, namely those with at least two target emotions. Because the comparison is made on increases rather than on raw posteriors, a hit reflects what steering changed rather than what the recognizer would have reported anyway. 
Rater proportions can produce exact ties for the dominant emotion. In this case, an utterance is scored as a hit when any tied dominant emotion also attains the largest posterior increase.

\noindent\textbf{S-SIM (Speaker similarity).} S-SIM is the cosine similarity between WavLM-base-sv~\citep{wavlm} speaker-verification embeddings of the generated utterance and its reference prompt. It detects the failure mode in which an injected direction alters the voice rather than the emotion.

\noindent\textbf{WER (Word error rate).}
Each generated utterance is transcribed using Whisper large-v3~\citep{whisper}, and the recognized words are compared with the prompted transcript.

\subsection{Subjective metrics}
\label{app:sub_metric}
Each of $K$ listeners assigns exactly one emotion from $\mathcal{E}$ to a clip. Write $c_k\in\mathcal{E}$
for the label given by listener $k$, and let
\begin{equation}
\hat p_{e}=\frac{1}{K}\sum_{k=1}^{K}\mathbf{1}\big[\,c_k=e\,\big]
\end{equation}
be the annotator distribution over the clip, which is non-negative and sums to one over $\mathcal{E}$.

\noindent\textbf{Dom-hit (Dominant emotion hit rate).}
The clip counts as a hit when the annotated label is the dominant target emotion,
\begin{equation}
\mathrm{dom\text{-}hit}=\mathbf{1}\Big[\ \arg\max_{e\in\mathcal{E}}\hat p_{e}\ =\
\arg\max_{e\in\mathcal{T}}p_{e}\ \Big],
\end{equation}
and dom-hit is the fraction of hits over clips. The listener argmax ranges over all of $\mathcal{E}$
rather than over $\mathcal{T}$, so a clip heard as an emotion that was never requested is a miss.

\noindent\textbf{Fidelity.}
Fidelity is one minus the total variation distance between the annotator and target mixtures,
\begin{equation}
\mathrm{Fidelity}=1-\tfrac{1}{2}\sum_{e\in\mathcal{E}}
\big|\,\hat p_{e}-p_{e}\,\big| ,
\end{equation}
which is one when the two distributions coincide and zero when their supports are disjoint.

\noindent\textbf{Naturalness preference.}
A trial presents one sentence rendered by two systems, $A$ and $B$, and returns one of three responses.
Scoring the response $r$ as
\begin{equation}
u(r)=
\begin{cases}
1, & r=\text{$A$ more natural},\\[2pt]
\tfrac{1}{2}, & r=\text{about the same},\\[2pt]
0, & r=\text{$B$ more natural},
\end{cases}
\end{equation}
the preference for $A$ is the mean of $u$ over observations, so that a panel with no systematic
preference scores $\tfrac{1}{2}$ regardless of how often it declines to choose.

%% file: appendix/exp_details.tex
\subsection{Emotion vector construction}
\label{app:vec_contruct}
We construct the emotion-vector library using the extraction procedure introduced by CoCoEmo~\citep{cocoemo}. We retain CoCoEmo's preprocessing, activation pooling, and steering layers (elaborated in Appendix~\ref{app:steer_layer}) to ensure that differences between CoCoEmo and EmoRES arise only from the steering rule.
The extraction library contains paired neutral and emotional utterances from ESD~\citep{esd}, CREMA-D~\citep{cremad}, and RAVDESS~\citep{ravdess}. Within each pair, the two recordings have the same speaker and linguistic content but differ in emotion. Each recording is checked using the emotion-recognition and signal-quality filters adopted from CoCoEmo. If either recording fails a filter, the complete pair is removed. IEMOCAP is excluded from vector construction and remains held out for evaluation.
For each frozen TTS backbone, we pass every retained recording through the model and cache the activation at the attention output of each selected steering layer. Let $\phi_\ell(x)$ denote the utterance-level activation obtained from recording $x$ at layer $\ell$ using CoCoEmo's pooling procedure. For emotion $e$, the corresponding emotional and neutral means are
\begin{equation}
  \bar h_{e}^{(\ell)}
  =
  \frac{1}{N_{e}}
  \sum_{i=1}^{N_{e}}
  \phi_\ell(x_{i,e}),
  \qquad
  \bar h_{\mathrm{neu},e}^{(\ell)}
  =
  \frac{1}{N_{e}}
  \sum_{i=1}^{N_{e}}
  \phi_\ell(x_{i,\mathrm{neu}}),
  \end{equation}
  where $(x_{i,e},x_{i,\mathrm{neu}})$ denotes a matched emotional and neutral pair with identical linguistic content spoken by the same speaker. 
   Because a pair is discarded whole whenever either side fails a filter, the surviving pairs differ across emotions, so $N_{e}$ is emotion-dependent and each emotion is contrasted against its own neutral partners, written $\bar h_{\mathrm{neu},e}^{(\ell)}$.
  The emotion vector at layer $\ell$ is then
  \begin{equation}
  v_{e}^{(\ell)}
  =
  \bar h_{e}^{(\ell)}
  -
  \bar h_{\mathrm{neu,e}}^{(\ell)}
  =
  \frac{1}{N_{e}}
  \sum_{i=1}^{N_{e}}
  \left[
  \phi_\ell(x_{i,e})
  -
  \phi_\ell(x_{i,\mathrm{neu}})
  \right],
  \end{equation}
a mean of within-pair differences. 
The symbol $\bar h_{\mathrm{neu}}$ in Section~\ref{section:method} denotes a common neutral
reference, and $\bar h_{e} = \bar h_{\mathrm{neu}} + v_{e}$ the corresponding neutral-referenced emotional mean, so that Eq.~\ref{eq:vemo} holds by construction. Both terms of the decomposition of the shared and residual components in Eq.~\ref{eq:split} are differences and are therefore invariant to that reference, and the emotion-dependent pairing does not introduce distinct neutral means into the shared component.

A separate vector is extracted for every emotion, selected layer, and TTS backbone because their activation spaces are not shared. Extraction is performed once with frozen model parameters and requires neither gradient computation nor weight updates. CoCoEmo uses these vectors directly. EmoRES decomposes the same vectors into shared and residual components without constructing a new vector library.

The emotion vectors are estimated from 8{,}050 paired emotional--neutral utterances, approximately 12.6 hours of speech counting both sides of every pair, drawn from the three public corpora listed above. 
EmoRES is training-free rather than label-free because it uses labeled emotional speech but requires no gradient-based adaptation.
For each backbone, extraction requires one inference pass over the corpus and produces a small set of backbone-specific vectors that is reused for all subsequent generations and evaluation corpora. 
In contrast, fine-tuning requires separate optimization and storage of an adapted checkpoint for each backbone.

\subsection{Steering layers}
\label{app:steer_layer}
In this work, to conduct controlled comparisons between CoCoEmo and our EmoRES method, we adopted CoCoEmo's configuration of steering layers 1, 6, and 8 for IndexTTS-2 and layers 14 and 17 for CosyVoice2.

\input{figures/cv2_layer_probe_tep}

\noindent\textbf{Probe accuracy against steering response.}
Figure~\ref{fig:cv2probetep} examines whether emotion separability at a layer is associated with its response to vector steering. For each layer, we compare the accuracy of logistic regression and nearest-centroid emotion probes with the TEP obtained when EmoRES is applied only at that layer. Probe accuracy is positively correlated with TEP across the 24 layers, with Spearman correlations of $0.673$ for logistic regression and $0.594$ for nearest centroid. The two layers with the highest probe accuracy are also the two highest-TEP admissible layers. These results suggest that emotion probes can provide a useful signal for identifying candidate steering layers, although probe accuracy and steering performance measure distinct properties.

\subsection{Input conditions}

Unless otherwise specified, every steered row in this paper receives two emotion signals. The backbone's native interface supplies the first, and vector steering adds the second on top of it. We follow Appendix E.1 of CoCoEmo~\citep{cocoemo} for both interfaces to ensure fair comparisons.

\noindent\textbf{Speaker prompt.} Unless otherwise specified, every arm clones its voice from a neutral reference recording. The reference is spoken by the target speaker but carries different content.
Neutrality ensures that emotion in the output originates in the model rather than in the prompt.

\noindent\textbf{Emotion-vector conditioning (IndexTTS-2).} IndexTTS-2 exposes a continuous emotion-vector interface, which we set from the requested emotion distribution following CoCoEmo's configuration. We leave that configuration unchanged, so conditioning is held constant across the composition rules we compare.

\noindent\textbf{Instruction conditioning (CosyVoice2).} CosyVoice2 uses natural-language instructions for emotion control. The instruction names the requested emotions in descending order of weight and omits their proportions: ``Say it in a happy tone.''
for a single target, and ``Say it in a blend of happy, sad, and angry emotions.'' for a mixture. We drop emotions of zero weight, because naming one at zero still requests it.

\subsection{Subjective evaluation}
\label{app:subsec-subj_detail}
For quality control, we included a small set of samples with unambiguous gold-standard labels and discarded all responses from annotators who failed these checks. We evaluated 360 IndexTTS-2 samples and 330 CosyVoice2 samples in both the emotion-labeling and naturalness-preference tasks. Each sample received at least nine independent annotations per task. Table~\ref{tab:subj} reports results computed from the retained annotations.

We estimate 95\% confidence intervals with bootstrap replicates at the sample level. For emotion labeling, the matched CoCoEmo and EmoRES outputs for each sample are resampled together, and the mean paired difference is recomputed for every resample. For naturalness, each comparison and all its retained judgments are resampled as one unit, after which the preference score is recomputed. The reported bounds are the 2.5th and 97.5th percentiles of the resulting bootstrap distributions.

%% file: figures/cv2_layer_probe_tep.tex
\begin{figure*}[th]
\centering
\includegraphics[width=\textwidth]{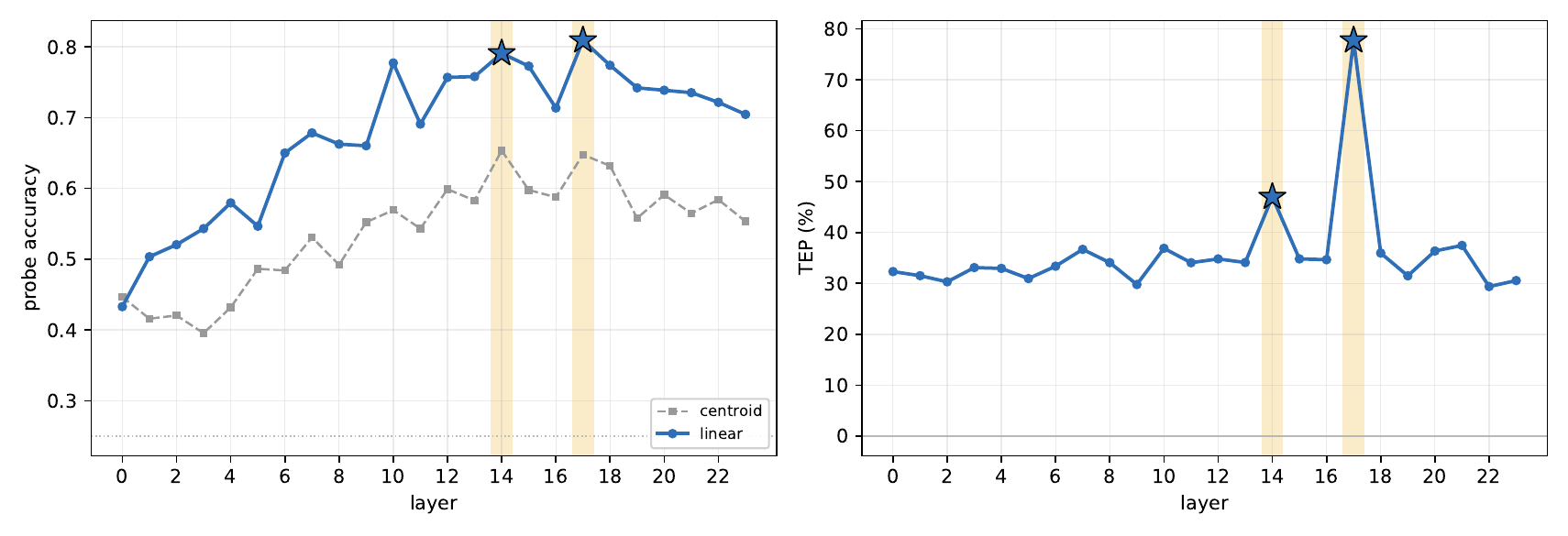}
\caption{Layer-wise emotion separability and causal steering performance for CosyVoice2 on CREMA-D. The left plot shows the accuracy of logistic regression and nearest-centroid emotion probes. The right plot shows TEP when EmoRES is applied to one layer at a time using $\lambda_c=1$, $\lambda_r=3$, and $\alpha=5$. Stars mark the two layers with the highest probe accuracy, while shading marks the two layers with the highest TEP.}
\label{fig:cv2probetep}
\end{figure*}

%% file: appendix/rand.tex
\input{tables/ablmix_rand}

Table~\ref{tab:ablmix} evaluates each steering component under three conditions: removed, replaced by a Gaussian control vector, or retained in its measured direction. Before injection, every nonzero steering vector is rescaled to a common reference norm given by the mean norm of the original emotion vector set. This reference is independent of $\lambda_c$ and $\lambda_r$, ensuring that differences between rows reflect component composition and direction rather than total steering magnitude. Consequently, the shared-only and residual-only rows are norm-matched endpoints of the same experimental design rather than lower-dose conditions.
Randomization is indicated directly in the coefficient columns. A coefficient written as $b\,\mathrm{rand}$ replaces the corresponding component with a Gaussian vector matched to that component's per-clip, per-layer norm and then scaled by $b$. For example, the condition $\lambda_c=1$ and $\lambda_r=``3\,\mathrm{rand}$'' is magnitude-matched to its unrandomized counterpart ($\lambda_c=1, \lambda_r=3$) and differs only in the residual direction. When both columns are marked $\mathrm{rand}$, the complete steering vector is replaced by a single Gaussian vector with the same final norm.

\subsection{Residual randomization}
Replacing the residual component with a norm-matched Gaussian vector substantially degrades all four emotion-control metrics. At $\lambda_r=1$, residual randomization reduces $\rho$ from 25.24\% to 5.62\% for IndexTTS-2 and from 32.48\% to 0.78\% for CosyVoice2. H-Rate decreases from 76.00\% to 69.51\% and from 78.92\% to 67.57\%, respectively, with corresponding reductions in E-SIM and TEP. The contrast becomes larger at $\lambda_r=3$. For IndexTTS-2, randomizing the residual reduces $\rho$ from 45.87\% to 0.58\% and H-Rate from 82.49\% to 67.57\%. For CosyVoice2, $\rho$ decreases from 51.94\% to 4.12\% and H-Rate from 84.76\% to 68.65\%. Because these pairs have the same residual allocation and final steering norm, the gains from residual enhancement cannot be reproduced by an arbitrary direction with the same magnitude.

\subsection{Shared component randomization}
The shared component also cannot be replaced by a norm-matched Gaussian. Across both residual strengths and backbones, shared randomization consistently reduces E-SIM and TEP. At $\lambda_r=3$, replacing the shared component lowers E-SIM from 64.80\% to 62.36\% and TEP from 39.98\% to 34.59\% for IndexTTS-2. For CosyVoice2, E-SIM decreases from 74.71\% to 71.09\% and TEP from 68.80\% to 61.63\%. The same replacement also reduces $\rho$ and H-Rate at $\lambda_r=3$ for both backbones.

\subsection{Complete randomization}
Replacing the complete steering vector with a norm-matched Gaussian vector fails to recover targeted emotion control. For IndexTTS-2, complete randomization produces a $\rho$ of $-0.50\%$ and an H-Rate of 67.14\%. For CosyVoice2, the corresponding values are $-0.94\%$ and 67.57\%. 
For both backbones, complete randomization yields E-SIM and TEP at or below the no-steer baseline, while its low $\rho$ and H-Rate show that the perturbation does not follow the requested emotion.
Together, the randomized controls demonstrate that the performance of EmoRES depends on the measured directions of both components. Neither the residual gain nor the shared contribution can be explained by steering magnitude, energy allocation, or an arbitrary perturbation of the hidden state.

%% file: tables/ablmix_rand.tex
\begin{table}[t]
\centering
\footnotesize
\setlength{\tabcolsep}{3pt}
\setlength{\belowcaptionskip}{5pt}
\caption{Component ablation on CREMA-D in the mixed-emotion setting with steering strength $\alpha = 5$. Every steered row is norm-matched to the same steering vector magnitude and differs from the others only in how that norm is split between the shared and residual components and in the direction each component points. A coefficient written ``$b\,\mathrm{rand}$'' marks a component replaced by a Gaussian at its own norm. Such rows are controls and are excluded from the bold comparison. Mean $\pm$ one standard deviation over five runs. Best per column and backbone among the unrandomized settings in bold. See Section~\ref{sec:app-ablmix} for the design and for the row contrasts.}
\label{tab:ablmix}
\resizebox{\textwidth}{!}{%
\begin{tabular}{llcccccccc}
\toprule
 & Method & $\lambda_c$ & $\lambda_r$ & E-SIM $\uparrow$ & TEP $\uparrow$ & $\rho$ $\uparrow$ & H-Rate $\uparrow$ & S-SIM $\uparrow$ & WER $\downarrow$ \\
\midrule
(a) & Ground Truth & -- & -- & 72.56 & 55.51 & -- & -- & 87.86 & 1.08 \\
\midrule
\multicolumn{10}{l}{\textbf{IndexTTS-2}} \\
(b) & No-steer & -- & -- & 55.00$\pm$0.36 & 16.83$\pm$1.39 & -- & -- & 87.18$\pm$0.16 & \textbf{6.30$\pm$0.05} \\
\addlinespace
(c) & Residual only & $0$ & $1$ & 63.05$\pm$0.29 & 35.88$\pm$0.77 & 45.16$\pm$5.92 & 82.16$\pm$2.06 & \textbf{87.64$\pm$0.18} & 6.36$\pm$0.06 \\
(d) & Shared randomized & $1\,\mathrm{rand}$ & $1$ & 59.55$\pm$0.42 & 28.50$\pm$1.21 & 21.52$\pm$5.27 & 74.49$\pm$2.04 & 87.74$\pm$0.15 & 6.62$\pm$0.09 \\
(e) & Shared \& Residual & $1$ & $1$ & 64.46$\pm$0.40 & 38.69$\pm$1.21 & 25.24$\pm$4.52 & 76.00$\pm$1.78 & 85.77$\pm$0.22 & 7.60$\pm$0.14 \\
(f) & Residual randomized & $1$ & $1\,\mathrm{rand}$ & 58.94$\pm$0.31 & 25.38$\pm$1.01 & 5.62$\pm$8.43 & 69.51$\pm$2.69 & 85.71$\pm$0.08 & 6.98$\pm$0.20 \\
(g) & Shared only & $1$ & $0$ & 59.51$\pm$0.33 & 26.64$\pm$1.18 & 5.30$\pm$7.86 & 69.62$\pm$2.21 & 85.13$\pm$0.22 & 6.69$\pm$0.26 \\
(h) & Both randomized & $\mathrm{rand}$ & $\mathrm{rand}$ & 54.42$\pm$0.39 & 15.65$\pm$1.64 & -0.50$\pm$10.42 & 67.14$\pm$3.45 & 87.62$\pm$0.18 & 6.60$\pm$0.19 \\
\addlinespace
(i) & Shared randomized & $1\,\mathrm{rand}$ & $3$ & 62.36$\pm$0.16 & 34.59$\pm$0.43 & 41.72$\pm$6.95 & 81.08$\pm$2.51 & 87.55$\pm$0.18 & 6.41$\pm$0.05 \\
(j) & Shared \& Residual & $1$ & $3$ & \textbf{64.80$\pm$0.25} & \textbf{39.98$\pm$0.88} & \textbf{45.87$\pm$6.99} & \textbf{82.49$\pm$2.34} & 86.90$\pm$0.21 & 6.78$\pm$0.16 \\
(k) & Residual randomized & $1$ & $3\,\mathrm{rand}$ & 57.00$\pm$0.23 & 21.50$\pm$0.92 & 0.58$\pm$8.96 & 67.57$\pm$3.08 & 86.70$\pm$0.19 & 6.92$\pm$0.17 \\
\midrule
\multicolumn{10}{l}{\textbf{CosyVoice2}} \\
(l) & No-steer & -- & -- & 59.70$\pm$0.36 & 34.01$\pm$0.96 & -- & -- & \textbf{87.36$\pm$0.21} & \textbf{0.53$\pm$0.06} \\
\addlinespace
(m) & Residual only & $0$ & $1$ & 72.21$\pm$0.27 & 63.82$\pm$0.56 & 39.94$\pm$5.04 & 80.54$\pm$1.53 & 87.21$\pm$0.16 & 1.14$\pm$0.15 \\
(n) & Shared randomized & $1\,\mathrm{rand}$ & $1$ & 67.65$\pm$0.21 & 53.83$\pm$0.86 & 19.14$\pm$4.74 & 73.41$\pm$1.17 & 87.22$\pm$0.14 & 0.70$\pm$0.08 \\
(o) & Shared \& Residual & $1$ & $1$ & 74.16$\pm$0.09 & 67.43$\pm$0.40 & 32.48$\pm$2.17 & 78.92$\pm$0.54 & 86.30$\pm$0.12 & 1.21$\pm$0.17 \\
(p) & Residual randomized & $1$ & $1\,\mathrm{rand}$ & 65.30$\pm$0.40 & 45.79$\pm$0.77 & 0.78$\pm$6.92 & 67.57$\pm$1.99 & 86.47$\pm$0.13 & 1.18$\pm$0.05 \\
(q) & Shared only & $1$ & $0$ & 65.92$\pm$0.26 & 47.53$\pm$0.68 & 7.20$\pm$5.03 & 70.16$\pm$2.14 & 85.77$\pm$0.13 & 2.30$\pm$0.19 \\
(r) & Both randomized & $\mathrm{rand}$ & $\mathrm{rand}$ & 59.20$\pm$0.35 & 32.34$\pm$1.01 & -0.94$\pm$2.98 & 67.57$\pm$1.15 & 87.46$\pm$0.17 & 0.62$\pm$0.07 \\
\addlinespace
(s) & Shared randomized & $1\,\mathrm{rand}$ & $3$ & 71.09$\pm$0.49 & 61.63$\pm$0.59 & 44.32$\pm$1.58 & 81.84$\pm$0.48 & 87.17$\pm$0.12 & 0.86$\pm$0.12 \\
(t) & Shared \& Residual & $1$ & $3$ & \textbf{74.71$\pm$0.24} & \textbf{68.80$\pm$0.56} & \textbf{51.94$\pm$6.30} & \textbf{84.76$\pm$2.07} & 86.73$\pm$0.18 & 1.30$\pm$0.10 \\
(u) & Residual randomized & $1$ & $3\,\mathrm{rand}$ & 63.41$\pm$0.24 & 42.35$\pm$1.37 & 4.12$\pm$4.68 & 68.65$\pm$1.48 & 86.85$\pm$0.07 & 0.92$\pm$0.09 \\
\bottomrule
\end{tabular}}
\end{table}

%% file: appendix/emoprompt_single.tex
Table~\ref{tab:emopromptone} examines whether EmoRES remains effective when the speaker prompt already expresses the requested emotion. 
A target-emotion prompt is a recording by the same speaker, on different content, that raters heard as the requested emotion.
Replacing the neutral prompt with a target-emotion prompt substantially improves the E-SIM and TEP of the no-steer condition for both backbones and datasets, confirming that the emotional content of the reference recording provides a useful control signal.

EmoRES nevertheless outperforms CoCoEmo in E-SIM and TEP for every backbone, dataset, and prompt type. With target-emotion prompts, EmoRES raises E-SIM from 75.66\% to 80.33\% and TEP from 66.40\% to 82.24\% for IndexTTS-2 on CREMA-D. On IEMOCAP, the corresponding improvements are from 79.36\% to 84.54\% and from 78.02\% to 89.17\%. The same pattern holds for CosyVoice2. EmoRES raises E-SIM from 82.10\% to 85.54\% and TEP from 84.54\% to 92.40\% on CREMA-D, and from 79.40\% to 82.82\% and from 80.04\% to 87.63\% on IEMOCAP. These results show that residual enhancement complements an emotional reference signal rather than merely compensating for a neutral prompt.

\input{tables/emoprompt_single}

%% file: tables/emoprompt_single.tex
\begin{table*}[h]
\centering
\setlength{\tabcolsep}{10pt}
\setlength{\belowcaptionskip}{5pt}
\caption{Single-emotion evaluation with neutral and \textbf{target-emotion speaker prompts}. All steered rows use $\alpha=5$ and $\lambda_c=1$. CoCoEmo is the $\lambda_r=1$ special case. Best among the synthesis rows is \textbf{bold},
second-best \underline{underlined}.}
\label{tab:emopromptone}
\begin{tabular}{lcccc}
\toprule
Method & E-SIM$\uparrow$ & TEP$\uparrow$ & S-SIM$\uparrow$ & WER$\downarrow$ \\
\midrule
\multicolumn{5}{c}{\textit{In-distribution evaluation on CREMA-D (dev)}} \\
\midrule
\multicolumn{5}{l}{\textbf{IndexTTS-2}} \\
\multicolumn{5}{l}{\quad\textit{neutral prompt}} \\
\quad No-steer & 59.21$\pm$0.38 & 30.16$\pm$1.37 & \textbf{88.63$\pm$0.16} & \textbf{6.54$\pm$0.10} \\
\quad CoCoEmo & \underline{68.08$\pm$0.45} & \underline{50.95$\pm$1.78} & 84.82$\pm$0.09 & \underline{6.73$\pm$0.12} \\
\quad EmoRES, $\lambda_r=3$ & \textbf{71.86$\pm$0.40} & \textbf{62.27$\pm$1.39} & \underline{85.09$\pm$0.22} & 8.47$\pm$0.47 \\\\
\multicolumn{5}{l}{\quad\textit{target-emotion prompt}} \\
\quad No-steer & 74.90$\pm$0.45 & \underline{70.89$\pm$1.17} & \textbf{86.21$\pm$0.26} & \textbf{7.25$\pm$0.11} \\
\quad CoCoEmo & \underline{75.66$\pm$0.24} & 66.40$\pm$1.31 & \underline{84.16$\pm$0.18} & \underline{8.42$\pm$0.20} \\
\quad EmoRES, $\lambda_r=3$ & \textbf{80.33$\pm$0.24} & \textbf{82.24$\pm$0.99} & 84.15$\pm$0.19 & 8.86$\pm$0.36 \\
\midrule
\multicolumn{5}{l}{\textbf{CosyVoice2}} \\
\multicolumn{5}{l}{\quad\textit{neutral prompt}} \\
\quad No-steer & 58.48$\pm$0.14 & 32.19$\pm$0.80 & \textbf{87.47$\pm$0.21} & \textbf{0.32$\pm$0.09} \\
\quad CoCoEmo & \underline{73.21$\pm$0.33} & \underline{68.60$\pm$0.78} & \underline{86.62$\pm$0.15} & \underline{0.68$\pm$0.07} \\
\quad EmoRES, $\lambda_r=3$ & \textbf{77.42$\pm$0.39} & \textbf{80.01$\pm$1.27} & 86.40$\pm$0.18 & 0.77$\pm$0.09 \\\\
\multicolumn{5}{l}{\quad\textit{target-emotion prompt}} \\
\quad No-steer & 69.48$\pm$0.33 & 60.93$\pm$0.61 & \textbf{86.46$\pm$0.16} & \textbf{0.16$\pm$0.09} \\
\quad CoCoEmo & \underline{82.10$\pm$0.17} & \underline{84.54$\pm$0.44} & \underline{85.85$\pm$0.24} & 0.68$\pm$0.07 \\
\quad EmoRES, $\lambda_r=3$ & \textbf{85.54$\pm$0.36} & \textbf{92.40$\pm$0.93} & 85.48$\pm$0.13 & \underline{0.67$\pm$0.13} \\
\midrule
\multicolumn{5}{c}{\textit{Out-of-distribution evaluation on IEMOCAP (OOD)}} \\
\midrule
\multicolumn{5}{l}{\textbf{IndexTTS-2}} \\
\multicolumn{5}{l}{\quad\textit{neutral prompt}} \\
\quad No-steer & 65.36$\pm$0.43 & 52.22$\pm$0.56 & \textbf{89.61$\pm$0.26} & \textbf{6.06$\pm$0.06} \\
\quad CoCoEmo & \underline{76.00$\pm$0.14} & \underline{71.10$\pm$1.21} & 83.48$\pm$0.27 & \underline{7.22$\pm$0.40} \\
\quad EmoRES, $\lambda_r=3$ & \textbf{81.45$\pm$0.45} & \textbf{85.33$\pm$1.09} & \underline{84.40$\pm$0.14} & 7.47$\pm$0.59 \\\\
\multicolumn{5}{l}{\quad\textit{target-emotion prompt}} \\
\quad No-steer & 76.09$\pm$0.21 & 75.86$\pm$0.44 & \textbf{88.99$\pm$0.32} & \textbf{6.61$\pm$0.16} \\
\quad CoCoEmo & \underline{79.36$\pm$0.24} & \underline{78.02$\pm$0.63} & 84.26$\pm$0.22 & \underline{8.25$\pm$0.42} \\
\quad EmoRES, $\lambda_r=3$ & \textbf{84.54$\pm$0.15} & \textbf{89.17$\pm$0.24} & \underline{84.96$\pm$0.38} & 8.52$\pm$0.32 \\
\midrule
\multicolumn{5}{l}{\textbf{CosyVoice2}} \\
\multicolumn{5}{l}{\quad\textit{neutral prompt}} \\
\quad No-steer & 62.85$\pm$0.17 & 47.35$\pm$0.96 & \textbf{88.06$\pm$0.23} & \textbf{2.41$\pm$0.25} \\
\quad CoCoEmo & \underline{73.49$\pm$0.32} & \underline{69.65$\pm$0.83} & \underline{87.43$\pm$0.30} & 3.07$\pm$0.14 \\
\quad EmoRES, $\lambda_r=3$ & \textbf{78.02$\pm$0.27} & \textbf{79.79$\pm$0.67} & 87.37$\pm$0.24 & \underline{3.06$\pm$0.15} \\\\
\multicolumn{5}{l}{\quad\textit{target-emotion prompt}} \\
\quad No-steer & 70.08$\pm$0.42 & 62.74$\pm$1.03 & \textbf{87.77$\pm$0.19} & \textbf{2.26$\pm$0.09} \\
\quad CoCoEmo & \underline{79.40$\pm$0.28} & \underline{80.04$\pm$0.85} & 87.40$\pm$0.10 & \underline{2.81$\pm$0.09} \\
\quad EmoRES, $\lambda_r=3$ & \textbf{82.82$\pm$0.33} & \textbf{87.63$\pm$0.37} & \underline{87.41$\pm$0.20} & 2.95$\pm$0.24 \\
\bottomrule
\end{tabular}
\end{table*}

%% file: appendix/nocond.tex
Table~\ref{tab:mixed_nocond} evaluates whether residual enhancement depends on the native emotion-conditioning interface of each backbone. Without emotion embeddings or instructions, the no-steer systems obtain substantially lower TEP and E-SIM than the steered systems. Both CoCoEmo and EmoRES improve these
metrics, demonstrating that the extracted directions can provide emotion control without assistance from native conditioning.

EmoRES improves E-SIM, TEP, $\rho$, and H-Rate over CoCoEmo for both backbones and datasets. On CREMA-D, $\rho$ increases from 11.43\% to 32.37\% for IndexTTS-2 and from 13.93\% to 44.62\% for CosyVoice2, while H-Rate increases from 70.92\% to 78.27\% and from 72.11\% to 82.05\%, respectively. On IEMOCAP, $\rho$ increases from 23.14\% to 44.76\% for IndexTTS-2 and from 22.08\% to 45.45\% for CosyVoice2, while H-Rate increases from 62.99\% to 73.65\% and from 61.98\% to 73.39\%. These results show that residual enhancement improves alignment with the requested emotion mixture even when steering provides the only explicit emotion control.

\input{tables/main_results_nocond}

%% file: tables/main_results_nocond.tex
\begin{table*}[!htbp]
\centering
\small
\setlength{\tabcolsep}{4pt}
\caption{Mixed-emotion evaluation \textbf{without native emotion conditioning}: IndexTTS-2 runs with no emotion embeddings and CosyVoice2 without the instructions, so the injected steering vectors are the only explicit emotion-control signal. CoCoEmo is the $\lambda_r=1$ special case. Mean $\pm$ standard deviation over five replicates. Best among the synthesis rows is \textbf{bold},
second-best \underline{underlined}.}
\label{tab:mixed_nocond}
\begin{tabular}{lcccccc}
\toprule
Method & E-SIM$\uparrow$ & TEP$\uparrow$ & $\rho\uparrow$ & H-Rate$\uparrow$ & S-SIM$\uparrow$ & WER$\downarrow$ \\
\midrule
\multicolumn{7}{c}{\textit{In-distribution evaluation on CREMA-D (dev)}} \\
\midrule
Ground Truth & 72.56 & 55.51 & -- & -- & 87.86 & 1.08 \\
\midrule
\multicolumn{7}{l}{\textbf{IndexTTS-2}} \\
No-steer & 47.84$\pm$0.10 & 2.58$\pm$0.27 & -- & -- & 89.38$\pm$0.16 & \underline{6.44$\pm$0.02} \\
CoCoEmo & \underline{52.77$\pm$0.29} & \underline{12.87$\pm$0.95} & \underline{11.43$\pm$5.87} & \underline{70.92$\pm$1.93} & \underline{89.85$\pm$0.14} & \textbf{6.32$\pm$0.08} \\
EmoRES, $\lambda_r=3$ & \textbf{56.18$\pm$0.23} & \textbf{20.51$\pm$0.67} & \textbf{32.37$\pm$6.39} & \textbf{78.27$\pm$2.37} & \textbf{90.12$\pm$0.13} & 7.13$\pm$0.10 \\
\midrule
\multicolumn{7}{l}{\textbf{CosyVoice2}} \\
No-steer & 48.80$\pm$0.30 & 4.69$\pm$0.65 & -- & -- & \textbf{89.71$\pm$0.16} & \textbf{1.04$\pm$0.48} \\
CoCoEmo & \underline{57.54$\pm$0.41} & \underline{25.71$\pm$1.82} & \underline{13.93$\pm$3.48} & \underline{72.11$\pm$1.69} & \underline{89.45$\pm$0.06} & \underline{1.51$\pm$0.40} \\
EmoRES, $\lambda_r=3$ & \textbf{67.83$\pm$0.54} & \textbf{55.35$\pm$0.69} & \textbf{44.62$\pm$7.03} & \textbf{82.05$\pm$2.52} & 88.03$\pm$0.28 & 1.59$\pm$0.53 \\
\midrule
\multicolumn{7}{c}{\textit{Out-of-distribution evaluation on IEMOCAP (OOD)}} \\
\midrule
Ground Truth & 60.68 & 39.82 & -- & -- & 87.06 & 11.95 \\
\midrule
\multicolumn{7}{l}{\textbf{IndexTTS-2}} \\
No-steer & 52.48$\pm$0.15 & 20.39$\pm$0.49 & -- & -- & \textbf{90.21$\pm$0.06} & \textbf{6.04$\pm$0.23} \\
CoCoEmo & \underline{56.14$\pm$0.22} & \underline{26.57$\pm$0.21} & \underline{23.14$\pm$4.94} & \underline{62.99$\pm$2.37} & 89.83$\pm$0.21 & \underline{6.12$\pm$0.25} \\
EmoRES, $\lambda_r=3$ & \textbf{58.88$\pm$0.22} & \textbf{29.95$\pm$0.18} & \textbf{44.76$\pm$3.38} & \textbf{73.65$\pm$1.83} & \underline{89.91$\pm$0.10} & 6.52$\pm$0.38 \\
\midrule
\multicolumn{7}{l}{\textbf{CosyVoice2}} \\
No-steer & 51.53$\pm$0.19 & 20.88$\pm$0.28 & -- & -- & \textbf{89.54$\pm$0.16} & \underline{9.56$\pm$0.41} \\
CoCoEmo & \underline{55.16$\pm$0.09} & \underline{26.98$\pm$0.24} & \underline{22.08$\pm$2.71} & \underline{61.98$\pm$1.29} & \underline{89.12$\pm$0.23} & \textbf{9.22$\pm$0.93} \\
EmoRES, $\lambda_r=3$ & \textbf{59.92$\pm$0.33} & \textbf{33.72$\pm$0.41} & \textbf{45.45$\pm$1.19} & \textbf{73.39$\pm$0.50} & 88.70$\pm$0.21 & 9.72$\pm$0.57 \\
\bottomrule
\end{tabular}
\end{table*}

%% file: appendix/threejudge_iemocap.tex
\input{tables/threejudge_iemocap}

Table~\ref{tab:threejudge} reports the held-out IEMOCAP evaluation scored by three different speech emotion recognizers: emotion2vec+ large~\citep{emotion2vec}, which is the recognizer used throughout the paper, Qwen3-Omni~\citep{qwen3-omni}, and the Odyssey~2024 WavLM~\citep{goncalves24_odyssey} model.
The comparison holds everything fixed except the recognizer. We reuse the audio already generated for Table~\ref{tab:mixed}, so the arms, the steering coefficients, the speaker prompts and the five prompt-resample replicates are identical.

Across both backbones, EmoRES outperforms CoCoEmo on TEP, $\rho$, and H-Rate under every recognizer, yielding improvements in all 18 comparisons. 
Despite their different absolute score ranges, all three recognizers produce the same ordering between CoCoEmo and EmoRES.
Because the generated audio and all generation conditions are unchanged across recognizers, this agreement indicates that the improvements in emotional control are not specific to emotion2vec+ large.

%% file: tables/threejudge_iemocap.tex
\begin{table}[t]
\centering
\small
\setlength{\tabcolsep}{10pt}
\caption{Mixed-emotion evaluation on IEMOCAP (OOD) under three different speech emotion recognizers. All steered rows use $\lambda_c=1$. Values are percentages, mean $\pm$ standard deviation over five replicates. Recognizers differ in calibration, so absolute levels are not comparable across blocks. Best values per block are marked \textbf{bold}.}
\label{tab:threejudge}
\begin{tabular}{llccc}
\toprule
SER judge & Method & TEP$\uparrow$ & $\rho\uparrow$ & H-Rate$\uparrow$ \\
\midrule
\multicolumn{5}{c}{\textit{IndexTTS-2}} \\
\cmidrule(lr){1-5}
\multirow{3}{*}{emotion2vec+ large} & No-steer & 30.76$\pm$0.44 & -- & -- \\
 & CoCoEmo & 36.22$\pm$0.80 & 22.00$\pm$4.83 & 64.34$\pm$2.39 \\
 & EmoRES, $\lambda_r=3$ & \textbf{41.11$\pm$0.39} & \textbf{48.13$\pm$2.44} & \textbf{77.29$\pm$1.67} \\
\cmidrule(lr){1-5}
\multirow{3}{*}{Qwen3-Omni} & No-steer & 37.08$\pm$0.22 & -- & -- \\
 & CoCoEmo  & 39.79$\pm$0.43 & 31.60$\pm$3.24 & 66.33$\pm$1.63 \\
 & EmoRES, $\lambda_r=3$ & \textbf{40.70$\pm$0.17} & \textbf{48.95$\pm$0.71} & \textbf{74.88$\pm$0.25} \\
\cmidrule(lr){1-5}
\multirow{3}{*}{Odyssey-WavLM} & No-steer & 27.92$\pm$0.08 & -- & -- \\
 & CoCoEmo  & 30.82$\pm$0.23 & 18.51$\pm$4.10 & 59.87$\pm$2.05 \\
 & EmoRES, $\lambda_r=3$ & \textbf{34.14$\pm$0.15} & \textbf{46.11$\pm$1.56} & \textbf{73.52$\pm$0.72} \\
\midrule
\multicolumn{5}{c}{\textit{CosyVoice2}} \\
\cmidrule(lr){1-5}
\multirow{3}{*}{emotion2vec+ large} & No-steer & 28.92$\pm$0.17 & -- & -- \\
 & CoCoEmo  & 37.44$\pm$0.36 & 39.13$\pm$1.56 & 70.90$\pm$0.81 \\
 & EmoRES, $\lambda_r=3$ & \textbf{40.12$\pm$0.29} & \textbf{52.10$\pm$1.05} & \textbf{77.82$\pm$0.68} \\
\cmidrule(lr){1-5}
\multirow{3}{*}{Qwen3-Omni} & No-steer & 32.69$\pm$0.18 & -- & -- \\
 & CoCoEmo  & 37.72$\pm$0.12 & 31.96$\pm$1.21 & 66.35$\pm$0.62 \\
 & EmoRES, $\lambda_r=3$ & \textbf{37.95$\pm$0.24} & \textbf{35.20$\pm$1.59} & \textbf{68.16$\pm$0.80} \\
\cmidrule(lr){1-5}
\multirow{3}{*}{Odyssey-WavLM} & No-steer & 24.58$\pm$0.08 & -- & -- \\
 & CoCoEmo  & 25.62$\pm$0.15 & 14.47$\pm$0.86 & 57.97$\pm$0.45 \\
 & EmoRES, $\lambda_r=3$ & \textbf{27.23$\pm$0.15} & \textbf{23.66$\pm$2.23} & \textbf{62.35$\pm$1.09} \\
\bottomrule
\end{tabular}
\end{table}